\documentclass[nonacm, sigconf]{nimeart}
\AtBeginDocument{%
  }
\setcopyright{cc}
\copyrightyear{2025}
\acmYear{2025}
\acmDOI{}

\renewcommand\footnotetextcopyrightpermission[1]{}
\acmISBN{}

\begin{document}

\title{Coupling the Heart to Musical Machines}

\author{Eric Easthope}
\email{eric@sequent.audio}
\affiliation{%
  \institution{University of British Columbia}
  \city{Vancouver}
  \state{British Columbia}
  \country{Canada}
}

\renewcommand{\shortauthors}{Eric Easthope}

\begin{abstract}
  Biofeedback is being used more recently as a general control paradigm for human-computer interfaces (HCIs).
  While biofeedback especially from breath has seen increasing uptake as a controller for novel musical interfaces, new interfaces for musical expression (NIMEs), the community has not given as much attention to the heart.
  The heart is just as intimate a part of music as breath and it is argued that the heart determines our perception of time and so indirectly our perception of music.
  Inspired by this I demonstrate a photoplethysmogram (PPG)-based NIME controller using heart rate as a 1D control parameter to transform the qualities of sounds in real-time over a Bluetooth wireless HCI.
  I apply time scaling to "warp" audio buffers inbound to the sound card, and play these transformed audio buffers back to the listener wearing the PPG sensor, creating a hypothetical perceptual biofeedback loop: changes in sound change heart rate to change PPG measurements to change sound.
  I discuss how a sound-heart-PPG biofeedback loop possibly affords greater control and/or variety of movements with a 1D controller, how controlling the space and/or time scale of sound playback with biofeedback makes for possibilities in performance ambience, and I briefly discuss generative latent spaces as a possible way to extend a 1D PPG control space.
\end{abstract}

\keywords{photoplethysmogram, heartbeat, sound, time, control}

\maketitle

\section{Introduction}
\subsection{Background}
While biofeedback has seen increasing uptake in the New Interface for Musical Expression (NIME) community, with breathing and markedly the performer's breath being a key biofeedback marker, somehow the community has almost entirely overlooked the heart.
The heart is just as intimate a part of music as breath is.
Some even argue that our seemingly innate sense of rhythm, and with rhythm our sense of cycles, and with cycles our sense of \textit{time}---even---emerges from interaction with our heartbeat \cite{Khoshnoud2024}.
Heart rate affected time judgement making it inconsequential to the music performance context; the heartbeat sets the clock rate for our perception of music.
Deliberate breathing control in the presence of music with variable BPM also had a direct effect on heart rate \cite{Watanabe2015} suggesting potential biofeedback loops between the sympathetic nervous system and auditory perception, which is now known to be intimately related to the primary motor system too \cite{Bedford2025}.

\subsection{Related Work}
Recent research has uncovered evidence that links our sense of rhythm to our heartbeat \cite{Khoshnoud2024}, suggesting the heart plays a role in our perception of time and rhythm.
Khoshnoud et al. showed that how we perceive time closely tied to interoceptive signals including our heartbeat.
They found a direct connection between the brain's processing of heartbeat signals and judgment of time duration from heartbeat-evoked potential (HEP) amplitude changes during the encoding phase of a timing task, suggesting that our internal sense of time is intimately connected to the rhythmic beating of our hearts.

\subsection{Objective}
Previous NIMEs \textit{have} recently explored breath and the detection of breathing changes as a control paradigm for NIMEs (Zhu et al., ``Fulgura Frango: Breath---Extended Harpsichord,'' \textit{NIME '24}).
It makes sense, given the direct physiological connections between breath and heartbeat, that if breath is a feasible NIME controller then heartbeat or at least derivations of other physiological quantities from heartbeat are also feasible NIME controllers.
For instance, because of the observations made about heart rate and judgement of time interval \cite{Khoshnoud2024}, and knowing that increasing BPM and increasing breath together increases heart rate, we think it could be possible to create a biofeedback-based control loop with a combination of real-time auditory BPM and breathing changes.

This paper seeks to introduce the heart as a NIME controller with simplicity in mind, hoping that other authors might find the incremental contribution and proof-of-concept more readily useful and easier to iterate upon than a more comprehensive and/or systematic accounting of PPG as a NIME paradigm as it covers but also goes beyond the heart.
This is in the same spirit as ``artifact''-based research or ``research-by-design.''
To this end we introduce a basic method to sample heart rate from a market-available PPG sensor using only open source libraries and show how this data connects to control a simple 1D sound parameter in real-time offering only a few suggestions about future work and leaving certain design constraints open-ended and unopinionated.

\section{Material \& Methods}
We took a Polar Verity Sense fitness tracker containing a consumer-grade photoplethysmogram (PPG) sensor, which supports 55 Hz wireless PPG on the wrist, arm, leg, or ankle with Bluetooth low-energy, and interfaced with its non-proprietary libraries using \texttt{bleak} and \texttt{asyncio} (Python).
Querying the heart rate universally unique identifier (UUID) as frequently as possible and tracking the timestamps of current heart rate values gave us a map of instantaneous heart rate with non-uniform spacing between samples due to the asynchronicity of sampling and small variations in query and processing time.
We linearly interpolated heart rate values to produce a uniformly sampled function of instantaneous heart rate.
This is directly mapped to some 1D control parameter.

This final mapping is not new but there is some choice in how to effectively use PPG.
For instance, changes in PPG compared to other modalities like electromyography (EMG) are relatively slow.

Following the recent evidence of heart rate directly affecting our time perceptions \cite{Khoshnoud2024}, one of the more obvious musical properties to map PPG to was time itself.
We could do this---varying the perception of musical time based on heart rate---by linearly ``warping'' the beats per minute (BPM) of music in real-time directly from changing PPG values.
Audio was streamed in its own thread using \texttt{PyAudio} (Python) at 44.1 kHz.
PPG values sampled from a paired Verity Sense tracker were processed concurrently based on their peaks to derive heart rate.
As buffered audio played, the length of samples to be read into the buffer was varied based on a \texttt{tempo} multiplier.
If there were more samples to be read into the buffer, more samples were played, producing an perceptual increase in tempo.
If there were \textit{less} samples read into the buffer, zeros were interspersed between buffered values, slowing down playback and producing an perceptual decrease in tempo.
This effectively ``warped'' the tempo of audio playback in real-time in response to PPG.

For demonstration \texttt{tempo} was set to 1.25, 25\% faster than normal playback speed, and variations in \texttt{tempo} were directly based on instantaneous heart rate values by multiplying the tempo based on the current heart rate value normalized to the previous few seconds of heart rate samples.
Changes to $\texttt{tempo}$ were clipped to $[1, 1.5]$ and changes in heart rate were computed in real-time (55 Hz) as more or less audio samples were buffered based on $\texttt{tempo}$ to produce bounded perceptual changes in tempo based on incoming PPG.

All Bluetooth connectivity to the Verity Sense tracker and real-time processing of PPG was done using Python on an Apple M1 chip with 16 GB of RAM.

\section{Result}
\begin{figure}[h]
  \centering
  \includegraphics[width=\linewidth]{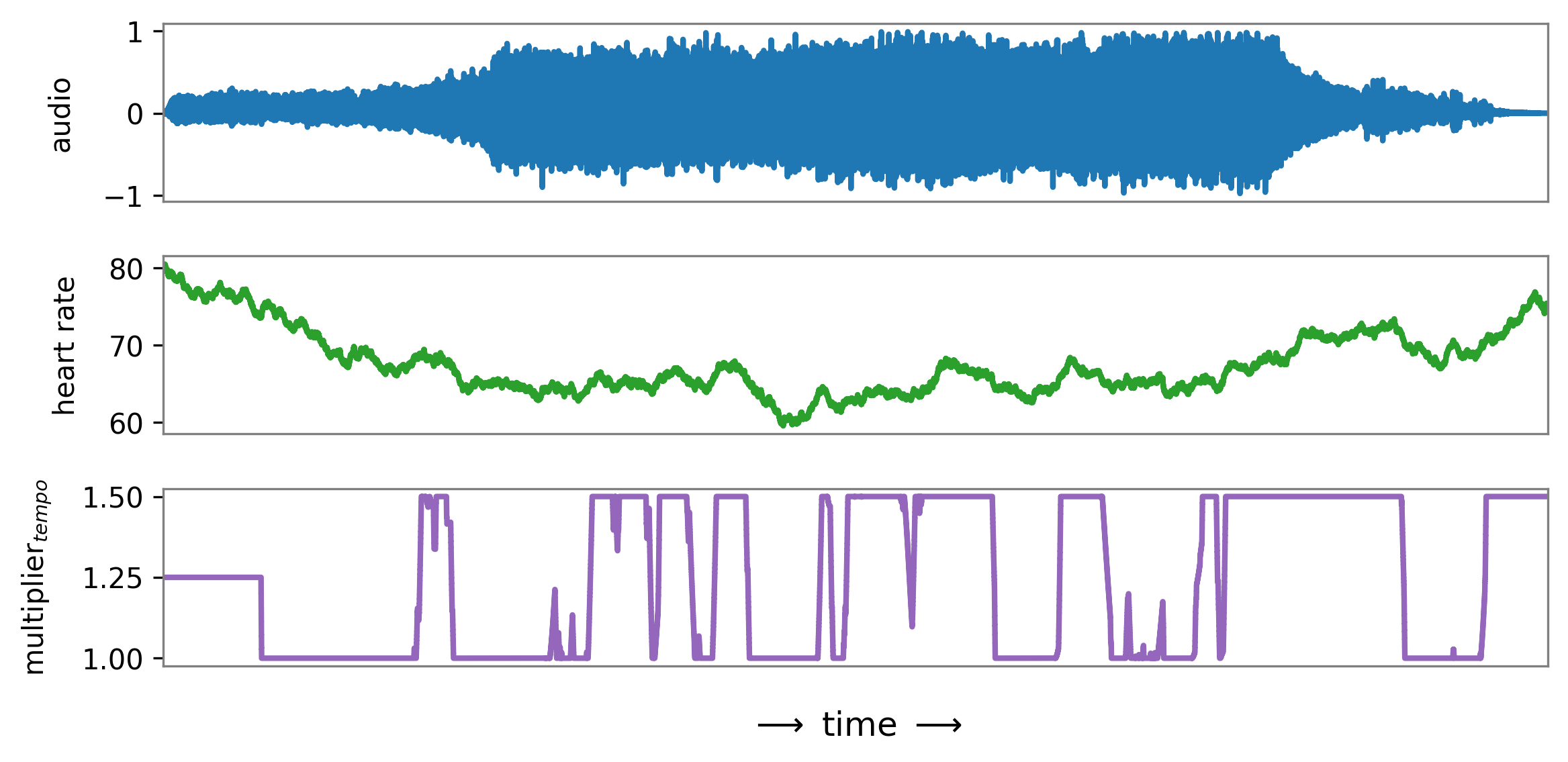}
  \caption{Audio amplitudes over time for a 5 minute, 38 second song sampled at 44.1 kHz from a WAV file with concurrent (simulated) heart rate data sampled from PPG and a derived ``multiplier'' for the real-time tempo of the playing audio.}
  \Description{Three stacked plots: a plot of music audio amplitudes, a plot of simultaneously derived heart rate data from PPG, and a plot of the derived multiplier value.}
  \label{fig:1}
\end{figure}

Figure \ref{fig:1} shows three stacked plots of values over time: a plot of music audio amplitudes for a 5 minute and 38 second song sampled at 44.1 kHz and loaded from a WAV file, a plot of simultaneously derived (simulated) heart rate data sampled from PPG, and a plot of the derived ``multiplier'' to change the real-time tempo of the playing audio.

\section{Discussion}
\subsection{Sound-heart-PPG biofeedback loops}
If we are able to create a sound-heart-PPG biofeedback loop with a PPG wearer perceiving and perhaps breathing also in response to music being transformed around them, it is possible that the creation of such a loop and the combination of a controller like this with other controllers such as the breath-based controller mentioned (Zhu et al., ``Fulgura Frango: Breath---Extended Harpsichord,'' \textit{NIME '24}) might afford greater control and/or a greater variety of movements than what is ordinarily possible with a 1D linear controller akin to a potentiometer or ``slider.''

\subsection{``Wandering'' controllers with slow biofeedback}
Many biofeedback signals are rather slow compared to the electromagnetic impulses that enable us to quickly control other NIMEs.
The PPG-based controller presented here depends on a gradual and continuous ``wandering'' of heart rate over time to trigger tempo changes.
But this offers an interesting basis for design if we are able to combine the gradual changes created by a PPG-based controller with other controllers that are more dynamic in short time.
That is to say that ``slow'' controllers---especially in the context of performance ambience and gradual changes in context, controlled changes in the scale of space and \textit{time} around music performance---might be just as essential.

\subsection{Extending ``slow'' biofeedback control with machine learning, non-linear maps}
Machine learning and specifically the application of generative latent spaces in their capacity to non-linearly map lower-dimension inputs, even 1D inputs, to sometimes significantly higher-dimensional outputs, makes them a viable candidate to extend the relatively ``slow'' 1D PPG-based control available to us to higher-dimensional controls and even controls that can leverage generative models trained on PPG data to transform the 1D PPG-based control space to other biofeedback and movement markers.

\section{Conclusion}
I demonstrated one of the first instances---to my knowledge, at least in the NIME community---of a PPG-based NIME controller using heart rate as a 1D control parameter to transform sounds in real-time.
I applied transformations to warp the temporal length of audio being played by the sound card in real-time using a wireless Bluetooth-based HCI.
As mentioned, playing a transformed audio buffer back to a PPG-wearing listener has the potential to create a perceptual biofeedback loop: changes in sound change heart rate to change PPG measurements to change sound.
I briefly discussed how a sound-heart-PPG biofeedback loop possibly affords greater control and/or variety of movements with a 1D controller, how controlling the space and/or time scale of sound playback with biofeedback makes for possibilities in performance ambience, and finally, how generative latent spaces are a possible way to extend a 1D PPG control space.

\bibliographystyle{ACM-Reference-Format}
\bibliography{paper}

\end{document}